\documentclass[useAMS,usenatbib]{mn2e}
\usepackage{graphicx,amsmath}

\title[Accelerated expansion induced by doubly-charged DM]{Accelerated expansion induced by Dark Matter with two charges}
\author[Steen H. Hansen]{Steen H. Hansen\\
Dark Cosmology Centre, Niels Bohr Institute, University of Copenhagen,
  Jagtvej 155, 2100 Copenhagen, Denmark}
\begin{document}

\pagerange{\pageref{firstpage}--\pageref{lastpage}} 

\maketitle

\label{firstpage}

\begin{abstract}
 The accelerated expansion of the universe has been established
  through observations of supernovae, the growth of structure, and the
  cosmic microwave background. The most popular explanation is
  Einsteins cosmological constant, or dynamic variations hereof. A
  recent paper demonstrated that if dark matter particles are endowed
  with a repulsive force proportional to the internal velocity
  dispersion of galaxies, then the corresponding acceleration of the
  universe may follow that of a cosmological constant fairly
  closely. However, no such long-range force is known to exist. A
  concrete example of such a force is derived here, by equipping the
  dark matter particles with two new dark charges. This result lends
  support to the possibility that the current acceleration of the
  universe may be explained without the need for a cosmological
  constant.
\end{abstract}




\section{Introduction}

 \label{sec:intro}
The acceleration of the Universe was first observed in SN1a data
\citep{perlmutter,riess} and has since been confirmed by a range of
independent observations including growth of the large scale structure
and the cosmic microwave background
\citep{komatsu, percival, blake, larson,hicken, blake2}.
Recent observational data is sufficiently
abundant and precise, that one may start seeing indications that
possibly not even dynamic versions of the cosmological constant
provide a self-consistent explanation of all the data
\citep{2021arXiv210302117D,2021arXiv210502667T,2021arXiv210503421K}.
It is therefore relevant to investigate
alternatives to (dynamics versions of) the cosmological constant. A
large number of such models have been discussed over the years, see
e.g. references in \citep{2021arXiv210404642M}.  Most of these
proposals conform with the idea that \citep{2003RvMP...75..559P}
{\em ''$\dots$this cosmic repulsion is a gravitational effect of the negative
gravitational mass density, not a new force law.''} A recent paper,
however, challenged this view, by demonstrating that if a new
inverse square-distance force exists amongst the dark matter particles,
then the resulting acceleration of the Universe is consistent with the
acceleration that is induced by a cosmological constant
\citep{2021ApJ...910...98L}. In that paper
the temporal evolution came about by letting the force depend on the
internal velocity dispersion of the dark matter halos: since structure
formation proceeds bottom-up, then the later structures will be more
massive and have higher velocity dispersions \citep{2021ApJ...910...98L}.
The main problem is, however, that no such long-range force is known to
exist in nature.

This paper shows that such a force indeed may exist. Using established
tools from spacetime algebra it is shown, that if one equips the dark
matter particles with two new dark-charges, where the two forces are
attractive and repulsive, respectively (likely because the carriers of
the forces have even and odd spins) then the non-cancelled part of the
forces has exactly the form needed to generate an accelerated universe
without a cosmological constant.

\section{Angular momentum}

To introduce the practical tool needed, namely spacetime algebra (STA),
let us start by addressing
how angular momentum can be made Lorentz invariant.

Even though both time and space separately are changed under a Lorentz
transformation, then the time-space vector, $(ct,\vec r)$, is
a proper 4-dimensional vector and hence has invariant length under
a change of reference.
Similarly, the energy-momentum vector, $(\varepsilon/c,\vec p)$,
is a proper 4-vector.

This is starkly contrasted by the 3-dimensional angular momentum,
$\vec L = \vec r \times \vec p$, where $\vec r$ and $\vec p$ are the
3-dimensional distance and momentum. This object cannot be combined
with anything to make a proper 4-dimensional vector. Similarly, the
dynamic mass moment, $\vec N = ct \vec p -\varepsilon \vec x/c $,
cannot be combined with anything to make a proper 4-vector~\citep{landau}.
Instead, there exists a rank-2 antisymmetric tensor, $M_{\mu \nu}$,
which is
the combination of $\vec L$ and $\vec N$, which has nice transformation
properties under a Lorentz transformation.
The properties of $M_{\mu \nu}$
are made most
simple and explicit in the language of spacetime
algebra~\citep{hestenes1966,
  2003AmJPh..71..691H}, where the
relevant object is written as \begin{equation} {\bf M} = r \wedge p \, , \label{eq:Mrq} \end{equation}
where
$r$ and $p$ are proper 4-dimensional vectors.
The wedge-operator, $\wedge$, is the natural 4-dimensional
antisymmetric generalization of the 3-dimensional cross-product,
$\times$. 
A brief teaser to STA calculations will be given now, and then
the technical details will be discussed correctly in the next section.
The basis of STA calculations is to
acknowledge that all observables exist in 4-dimensional Minkowski
space.
In STA there is one
unique derivative, $\nabla$, which contains both time- and
space-derivatives, and hence the only natural equation to write
for ${\bf M}$ 
is
\begin{equation} \nabla {\bf M} = j_M \, , \label{eq:delMj}\end{equation} where $j_M$ is some source. Forces
appear in STA by contracting ${\bf M}$ with a proper 4-dimensional
vector, $w$, to get ${\bf M} \cdot w$, which for instance gives Newtons
gravitational law. The only thing missing is, that one needs
observations to determine the constant in front of the force, which in
this case is the gravitational constant, $G$.

\section{Spacetime algebra}
\label{sec:sta}

After this brief teaser, the details of the algebra can now
be introduced correctly.
Spacetime algebra (STA) starts with Minkowski space, ${\cal M}_{1,3}$,
using the metric signature $(+,-,-,-)$, and a chosen basis
$\{\gamma_\mu \}_{\mu=0}^3$ of ${\cal M}_{1,3}$.

These four orthonormal vectors constitute the basis for 1-blades.  The
six antisymmetric products $\gamma _{\mu \nu} \equiv \gamma_{\mu}
\gamma_{\nu} $ are called the 2-blades. The product is in general
given by the sum of the dot and wedge products: $a b = a \cdot b + a
\wedge b$ \citep{hestenes2015,doranlasenby}.  The four 3-blades,
$\gamma _{\mu \nu \delta}$, are given by $\gamma_{\mu \nu \delta} =
\gamma_\mu \gamma_\nu \gamma_\delta $.  Finally one reaches the
highest grade, the pseudoscalar $I \equiv \gamma _0 \gamma _1 \gamma
_2 \gamma _3$, which represents the unit 4-volume in any basis, with
the property that $I^2 = -1$.

The generalized angular momentum mentioned above is called a
bi-vector.  The bi-vectores are oriented plane segments, and examples
also include the electromagnetic field $ {\bf F}$
\citep{hestenes2015,doranlasenby}.  Vector-arrows are used above spatial
3-vectors like $\vec E$ or $\vec p$, no-vector-arrows are used for
proper 4-vectors like $r$ and $p$, and boldface is used for bi-vectors
like ${\bf F}$ and ${\bf M}$.

One of the reasons for the success of STA is that the derivative
$\nabla {\bf F} = \nabla \cdot {\bf F} + \nabla \wedge {\bf F}$
naturally contains both time and space derivation.  When choosing a
time-direction, $\gamma_0$, one can decompose the derivative along a
direction parallel to and perpendicular to $\gamma_0$, $\nabla =
\left( \partial _0 - \vec \nabla \right) \gamma_0 $, where $ \vec
\nabla $ is the frame-dependent relative 3-vector derivative.  This
choice of frame also allows one to connect a 4-vector $w$ with its
para-vector, $w_0+\vec w$, via $\gamma_0$~\citep{doranlasenby}, namely
$w = \left( w_0 + \vec w \right) \gamma_0$.  Furthermore, the
right-multiplication by the timelike vector $\gamma_0$ isolates the
relative quantities of that frame~\citep{dressel2015}, e.g. $r
\gamma_0 = \left( ct + \vec r \right)$.

\section{Electromagnetism in STA}

The case of electromagnetism in STA is well described in the
literature \citep{hestenes2015, dressel2015}.
The starting point may be taken with the bi-vector \citep{hansenEM}
\begin{equation} {\bf F} = \frac{q}{m} \, r \wedge p \, . \label{eq:Frq} \end{equation}
When comparing to eq.~(\ref{eq:Mrq}) one notes that the only
difference is the exchange of mass by charge.

The simplest non-trivial equation is in this case given by
\begin{equation}
  \nabla {\bf F} = j_e \, ,
  \label{eq:maxwell}
\end{equation}
where the current, $j_e$, only contains an electric part in the absense of
magnetic monopoles. By making the identification
\begin{equation}
{\bf F} = \vec E + \vec B \, I \, ,
\label{eq:FEB}
\end{equation}
it is straight forward to derive the 4 Maxwells equations
\citep{hestenes2015, dressel2015}, which include
\begin{eqnarray}
\vec  \nabla \cdot \vec E &=& \frac{\rho_e}{\varepsilon_0} \, ,\label{eq:gauss} \\
  \partial_0 \vec B + \vec \nabla \times \vec E &=& 0 \, .
  \end{eqnarray}

Whereas ${\bf F}$ is a proper geometric object of electromagnetism,
then the separation into $\vec E$ and $\vec B$ fields requires
specification of a frame by the choice of $\gamma_0$.  At this point
the connection with the tensor $F_{\mu \nu}$ can be made explicit by
noticing \citep{dressel2015}
\begin{eqnarray}
  {\bf F} &=& \left( {\bf F} \cdot \gamma_0 \right) \gamma_0 + \left( {\bf F} \wedge \gamma_0 \right) \gamma_0
  \nonumber \\
  &=&\vec E + \vec B \, I   
\\
  &=&
  E_1 \gamma _{10} +E_2 \gamma _{20} +E_3 \gamma _{30} \nonumber \\
  && + B_1 \gamma_{32} + B_2 \gamma_{13} + B_3 \gamma_{21} \, ,
\end{eqnarray}
where $E_i$ and $B_i$ are the components of the corresponding
3-vectors, and $\gamma_{\mu \nu}$ are the 2-blades as described in
section~\ref{sec:sta}.

It is important to emphasize, that eq.~(\ref{eq:maxwell}) is not
merely a matter of compact notation: it is instead the only logical
extension beyond the most trivial equation in STA, $\nabla {\bf F} =
0$ (which leads to the four Maxwells equations without sources). The
only thing remaining is to connect the bi-vector field to observables:
the radial dependence is found from Gauss' law, eq.~(\ref{eq:gauss}),
and the units of $\vec E$ and $\vec B$ are established through
measurements.

\section{Appearance of forces}
In STA forces are derived by contracting  the bi-vectors with a proper
velocity vector, $w$, e.g.  the Lorentz force is derived directly from
${\bf F} \cdot w $~\citep{dressel2015}
\begin{equation}
  \left(  {\bf F} \cdot w \right) q \frac{d\tau}{dt} \gamma_ 0=
  q \vec E \cdot \vec v +
  q \left(  \vec E + \vec v \times \vec B\right) \, ,
\label{eq:lorentzforce}
\end{equation}
where the first term on the r.h.s.~is the rate of work,
$d\varepsilon/d(ct)$, and the last parenthesis on the r.h.s.~is
the classical Lorentz force, where $\vec w = \gamma \vec v$.
One notes that the $\gamma$ here is the
relativistic Lorentz factor, and should not be confused with the bases
for Minkowski space, $\gamma_\mu$.

Similarly the gravitational force appears from the contraction
of ${\bf M}$ and a proper 4-velocity
\begin{equation}
  \left( {\bf M} \cdot w  \right) m \frac{d\tau}{dt} \gamma_0 = -m \vec N \cdot \frac{\vec v}{c} +
  m \left( -  \vec N - \frac{\vec v}{c} \times \vec L \right) \,.
  \label{eq:gravforce}
\end{equation}
The first term on the r.h.s. is similar to a rate of work, and the first
term in the parenthesis leads to Newtons gravitational force~\citep{hansenNEWTON}.

When one considers gravitational forces, then there is an extra detail,
which arises if the structure under consideration contains a dynamical
term proportional to the velocity dispersion, $\sigma^2$. This could
for instance arise in a galaxy cluster where the galaxies and dark
matter particles are orbiting in the local gravitational potential. In
this case the potential is minus 2 times the kinetic energy according
to the virial theorem~\citep{bt2}, $2T+U=0$, and hence one can write
the energy as
\begin{equation}
\varepsilon = mc^2 - \frac{1}{2} m \sigma^2 \, .
\label{eq:correction}
\end{equation}
The correction in eq.~(\ref{eq:correction}) above is the first order
correction. It is similar to the first order relativistic correction
to the kinetic energy of a gas, where the difference between the
relativistic and restenergies typically reads

\begin{equation}
  m c^2 \left(
\frac{1}{\sqrt{1 - \frac{v^2}{c^2}}} - 1 \right) = \frac{3}{2} k T \, .
\end{equation}

The fact that it is exactly this correction to the energy which is
relevant, comes from the dynamic mass moment, ${\bf N}$, as shown
in \cite{hansenNEWTON}.  It is the expectation that the exact same
expression can be derived from general relativity (GR) in a weak field
limit, as the natural correction to the energymomentum tensor when one
includes a structure with non-zero internal dispersion.

From this extra term there will be a correction to the normal
Newtonian force, which is proportional to the velocity dispersion
squared~\citep{hansenNEWTON}.  This correction can be of the order
$10^{-5}$ for a galaxy cluster of mass $M_{\rm cluster} = 10^{15}
M_\odot$, and of the order $10^{-9}$ for a dwarf galaxy of mass
$M_{\rm dwarf} = 10^{9} M_\odot$.

\section{A new repulsive force for dark matter particles}
\label{sec:repulsive}

Having seen how the Lorentz force of electromagnetism
and Newtons gravitational law appear naturally from the
bi-vectors ${\bf F}$  and ${\bf M}$, in eqs.~(\ref{eq:Frq}, \ref{eq:Mrq}),
it is straight forward to generalize to new forces. Let us
imagine that the dark matter particles are equipped with a new charge, $q_r$.
The index, $r$, refers to {\em repulsive}, and one imagines that a quantum
field theoretical description of the force-carrier
is that of a spin-1 particle, such that equal charges
repel each other. At this point let us consider an asymmetric
creation of dark matter, such that all dark matter particles
carry the same charge.

A new bi-vector is defined, ${\bf D}_r= \vec D_r + \vec C_r \, I$,
(compare with the case of electromagnetism in eqs.~(\ref{eq:FEB},
\ref{eq:Frq})) where the $D$ refers to the dark matter particle
\begin{equation} {\bf D}_r = \frac{q_r}{m} \, r \wedge p \, . \label{eq:Drrq}
\end{equation}
From the dynamical equation
\begin{equation} \nabla {\bf D}_r = j_d \, ,
\end{equation} where
$j_d$ is a source term, one gets 4 equations, of which one is $\nabla
\cdot {\bf D}_r = \rho_d / \varepsilon_d$, where $\rho_d$ is the
number density of the dark matter particle.  $\varepsilon_d$ is
  similar to the vacuum permittivity, $\varepsilon_0$, of
  electromagnetism.  Integrating this equation (just like Gauss' law)
one gets that the field $\vec D_r$ is inverse square-distance. The
corresponding forces are found from ${\bf D}_r \cdot w$, of which the
main term gives a force
\begin{eqnarray} \frac{dp }{d\tau}
\gamma _0 &=& q_r \, \left( {\bf D}_r \cdot w \right) \nonumber \\ &\approx&
q_r \gamma \vec D_r \, .  \end{eqnarray} The magnitude of the force must be
determined from observations, in the same way that one measures $G$ for
the gravitational force, and $\varepsilon_0$ for the Coulomb force.

There is one very important detail, namely the sign of the minor
correction in eq.~(\ref{eq:correction}). Since the structure was
already created through gravity, the correction will come with a
negative sign (as in eq.~(\ref{eq:correction})) because the particles
have repulsive forces: the particles try to push each other apart, and
are therefore in a slightly higher energy-state than they would
prefer.

\section{A new attractive force for the dark matter particles}
\label{sec:attractive}
The dark matter particle is now also equipped with a
second charge, $q_a$. The index $a$ indicates that this is an {\em
  attractive} force, and hence one imagines it is generated by an
even-spin force carrier. Everything repeats itself from section
\ref{sec:repulsive}, except for two details: that one is using
$q_a$ and $\varepsilon_a$, and that the minor correction from
eq.~(\ref{eq:correction}) will come with the opposite sign:
the particles are attracted to each other, and are hence in a lower
energy state by being close together.
This sign difference will be very important in the next section.
The force again turns out to be
inverse square-distance, however, it will be attractive. The magnitude
of the force, expressed through $\varepsilon_a$ to make the connection
with $\varepsilon_0$ of electromagnetism explicit, must be determined
from measurements.

\section{Two combined forces}

In the discussion above, there is nothing indicating the magnitude
of the forces: both were inverse distance-square forces, and their
magnitude could be virtually anything.

If each dark matter particle happens to be equipped with both charges,
and for some reason the magnitude of the two forces happen to be
identical, $q_a^2/\varepsilon_a = q_r^2/\varepsilon_r$, then the main
force terms 
will cancel, where the main terms correspond to the first terms on the
r.h.s.\ of eq.~(\ref{eq:correction}).
This is because the one force was (constructed to be) attractive and
the other repulsive, but with equal strengths.  However, the minor
corrections (the last term in eq.~(\ref{eq:correction})) have
opposite signs, and the combined force will therefore be non-zero.  The
resulting combined force is repulsive and of the form
\begin{equation}
\frac{q_r^2}{\varepsilon_r} \, \left( 1 - \frac{1}{2} \frac{\sigma^2}{c^2} \right)
 \, \frac{\hat r}{r^2} -
\frac{q_a^2}{\varepsilon_a} \, \left( 1 + \frac{1}{2} \frac{\sigma^2}{c^2} \right)
\, \frac{\hat r}{r^2}
=
\frac{q_a^2}{  \varepsilon_a} \frac{\sigma^2}{c^2} \frac{\hat r}{r^2} \, ,
\label{eq:generalforce}
\end{equation}
where the dispersion has been
normalized to the speed of light.

This form is exactly of the shape needed for the suggestion of
\cite{2021ApJ...910...98L} \begin{equation} \kappa \, \frac{\sigma^2}{c^2}
G m_1 m_2 \frac{\hat r}{r^2} \, .
\label{eq:k2}
\end{equation} By comparing the two expressions above, one sees that the
combination of the dark matter charge and their strength, must be
equal to the gravitational force times a constant $\kappa$, which
according to \cite{2021ApJ...910...98L} should be of the order
$\kappa \sim 10^6 - 10^8$.

\section{Observational constraints}
A standard comparison of the strength of forces gives that
the gravitational attraction between 2 protons is about
36 orders of magnitude smaller than the electromagnetic
repulsion. 
One notices that a force only 6-8 orders
of magnitude stronger than gravity is needed, which means 28-30 orders
of magnitude weaker than electromagnetism. Since the above
calculations were imagined in an asymmetric dark matter model
(all dark matter particles are created with the same charge),
then that can be imagined as a charge which is 14-15 orders
of magnitude smaller than the electric charge.

Since all main forces cancel
between the attractive and repulsive forces,
this implies that the
evolution of the universe will be entirely unaffected
by the charges themselves. The only difference is that in the late
universe, when the dispersions in galaxies starts being
significant, there will be an overall repulsive force,
which may accelerate the expansion of the universe.

Dark charges and the corresponding dark radiation have
been discussed for many years \citep{2009PhRvD..79b3519A}.
Impressively many models have been considered, some with massive dark
photons (which is not relevant for the model considered here), some
models with massless dark (or hidden) photons, with symmetric and
asymmetric dark matter, milicharged dark matter, and naturally
a very large number of models where there is some kind of
interaction between the dark and visible sectors. Most of
the constraints are highly model dependent and cannot be
reviewed here, instead the reader is referred to the references in
the reviews
\citep{2017arXiv170704591B, 2020arXiv200501515F}.

The calculations above are entirely classical, in the sense that STA
only allows one to derive Maxwells equations and the Lorentz
force. In order to perform a quantum field theoretical derivation of
the forces, one would have to define and calculate the relevant
Feynman diagrams.  It is the expectation that in the exact same way as
QED generalized classical electromagnetism, a similar quantum field
theoretical generalization would allow one to derive the dark forces
described above. In this connection it will be particularly
interesting how one at the quantum level will make the two different
forces (mediated by different spin particles) cancel. Naturally this
point is beyond the scope of the classical derivation presented
above.

Any quantum field contributes zero-point energy, which might add a
contribution to the cosmological constant which is many orders of
magnitude larger than observationally acceptable \citep{weinberg}.
This goes very much counter to the suggestion of this paper, namely to
avoid the need for a cosmological constant.  The new dark charges
proposed above are only likely to worsen this situation, and the
present paper does in no way attempt to solve this issue.

One significant concern mentioned in \cite{2021ApJ...910...98L}
is the stability of structures like galaxies and dwarf galaxies: if
the force would be a standard ``electromagnetic-like'' force, then
there would be large forces internally in a dwarf galaxy, which would
rip the structure apart. In the present model all these large internal
forces are exactly cancelled, and hence that concern is completely
avoided.  In order to test if this model is indeed able to explain the
observed accelating universe, it will be necessary to perform a
numerical simulation, where the acceleration is accurately calculated
in each time-step, from the actual velocity dispersions. Such a
calculation is very likely to result in a temporal evolution, which is
not an exact match to the evolution of a (dynamic versions of a)
cosmological constant.  The resulting observables can then be compared
with accurate data from supernovae, growth of structures, and the
cosmic microwave background, and in this way the models can be
compared, testet and potentially observationally rejected.

\section{Conclusion}

Using the framework of spacetime algebra it is shown that when
equipping the dark matter particles with two different charges
(attractive and repulsive, respectively) with identical strengths,
then the non-cancelled part of the long-range forces is proportional
to velocity dispersion squared.  This is an example of a force which
was recently suggested to be able to accelerate the expansion of the
universe without the need for a cosmological constant.  The magnitude
of the strength is about 30 orders of magnitude smaller than
electromagnetism, and hence perfectly allowed observationally.

\section*{Data availability}
No new data were generated or analysed in support of this research.


\label{lastpage}


\begin{thebibliography}{00}


\bibitem[\protect\citeauthoryear{Ackerman et al.}{2009}]{2009PhRvD..79b3519A} Ackerman, L., Buckley, M.~R., Carroll, S.~M., et al.\ 2009, PRD, 79, 023519. 

  
\bibitem[\protect\citeauthoryear{Battaglieri et al.}{2017}]{2017arXiv170704591B} Battaglieri, M., Belloni, A., Chou, A., et al.\ 2017, arXiv:1707.04591

  \bibitem[\protect\citeauthoryear{Binney \& Tremaine}{2008}]{bt2} Binney, J. \& Tremaine, S.\ 2008, Galactic Dynamics: Second Edition,
  Princeton University Press, 2008.

    \bibitem[\protect\citeauthoryear{Blake et al.}{2011}]{blake}   Blake, C. {\it et al.},   Mon.\ Not.\ Roy.\ Astron.\ Soc.\  {\bf 415} (2011) 2876
    
  \bibitem[\protect\citeauthoryear{Blake et al.}{2012}]{blake2}   Blake, C. {\it et al.},   arXiv:1108.2637 [astro-ph.CO].

    
    
\bibitem[\protect\citeauthoryear{Dainotti et al.}{2021}]{2021arXiv210302117D} Dainotti, M.~G., De Simone, B., Schiavone, T., et al.\ 2021, arXiv:2103.02117


  
  
  \bibitem[\protect\citeauthoryear{Dressel et al.}{2015}]{dressel2015} Dressel, J., Bliokh, K.~Y., \& Nori, F.\ 2015, Phys.Rep., 589, 1. 

  \bibitem[\protect\citeauthoryear{Doran \& Lasenby}{2007}]{doranlasenby} Doran, C. \& Lasenby, A.\ 2007, Geometric Algebra for Physicists,
  Cambridge University Press, 2007

\bibitem[\protect\citeauthoryear{Fabbrichesi et al.}{2020}]{2020arXiv200501515F} Fabbrichesi, M., Gabrielli, E., \& Lanfranchi, G.\ 2020, arXiv:2005.01515


  \bibitem[\protect\citeauthoryear{Hansen}{2021}]{hansenEM} Hansen, S. H.\ 2021, subm.

  \bibitem[\protect\citeauthoryear{Hansen}{2021}]{hansenNEWTON} Hansen, S. H.\ 2021,
    MNRAS, 506, L16. 


\bibitem[\protect\citeauthoryear{Hestenes}{1966}]{hestenes1966} Hestenes, D.\ 1966, ``Space-Time Algebra'',
  Gordon and Breach Science Publishers, New York, 1966
  
\bibitem[\protect\citeauthoryear{Hestenes}{2003}]{2003AmJPh..71..691H} Hestenes, D.\ 2003, American Journal of Physics, 71, 691. 


\bibitem[\protect\citeauthoryear{Hestenes}{2015}]{hestenes2015} Hestenes, D.\ 2015, ``Space-Time Algebra'',
  Springer International Publishing Switzerland 2015

  
\bibitem[\protect\citeauthoryear{Hicken et al.}{2009}]{hicken}   Hicken, M. {\it et al.},   Astrophys.\ J.\  {\bf 700}, 1097 (2009)
\bibitem[\protect\citeauthoryear{Komatsu et al.}{2011}]{komatsu}   Komatsu, E. {\it et al.}  [WMAP Collaboration],   Astrophys.\ J.\ Suppl.\  {\bf 192}, 18 (2011)

  \bibitem[\protect\citeauthoryear{Krolewski et al.}{2021}]{2021arXiv210503421K} Krolewski, A., Ferraro, S., \& White, M.\ 2021, arXiv:2105.03421

\bibitem[\protect\citeauthoryear{Landau \& Lifshitz}{1975}]{landau} Landau, L.~D. \& Lifshitz, E.~M.\ 1975, Course of theoretical physics,
   Pergamon Press, 1975

  \bibitem[\protect\citeauthoryear{Larson et al.}{2011}]{larson}   Larson, D. {\it et al.},   Astrophys.\ J.\ Suppl.\  {\bf 192}, 16 (2011)

  \bibitem[\protect\citeauthoryear{Loeve et al.}{2021}]{2021ApJ...910...98L} Loeve, K., Nielsen, K.~S., \& Hansen, S.~H.\ 2021, ApJ, 910, 98.  

\bibitem[\protect\citeauthoryear{Motta et al.}{2021}]{2021arXiv210404642M} Motta, V., Garc{\'\i}a-Aspeitia, M.~A., Hern{\'a}ndez-Almada, A., et al.\ 2021, arXiv:2104.04642
    
\bibitem[\protect\citeauthoryear{Peebles \& Ratra}{2003}]{2003RvMP...75..559P} Peebles, P.~J. \& Ratra, B.\ 2003, Reviews of Modern Physics, 75, 559. 

\bibitem[\protect\citeauthoryear{Percival et al.}{2010}]{percival}   Percival, W.~J. {\it et al.}  [SDSS Collaboration],   Mon.\ Not.\ Roy.\ Astron.\ Soc.\  {\bf 401}, 2148 (2010)
\bibitem[\protect\citeauthoryear{Perlmutter et al.}{1999}]{perlmutter}   Perlmutter, S. {\it et al.}  [Supernova Cosmology Project Collaboration],   Astrophys.\ J.\  {\bf 517}, 565 (1999)

\bibitem[\protect\citeauthoryear{Riess et al.}{1998}]{riess}   Riess, A.~G. {\it et al.}  [Supernova Search Team Collaboration],   Astron.\ J.\  {\bf 116}, 1009 (1998)


    \bibitem[\protect\citeauthoryear{Teng et al.}{2021}]{2021arXiv210502667T} Teng, Y.-P., Lee, W., \& Ng, K.-W.\ 2021, arXiv:2105.02667


    \bibitem[\protect\citeauthoryear{Weinberg}{1989}]{weinberg}
      Weinberg, S.\ 1989, Reviews of Modern Physics, 61, 1. 

 

\end{thebibliography}
\end{document}